\documentclass[conference]{IEEEtran}
\IEEEoverridecommandlockouts
\ifCLASSINFOpdf
  \usepackage[pdftex]{graphicx}
\else
  \usepackage[dvips]{graphicx}
\fi

\usepackage{algorithm}
\usepackage{algpseudocode}
\usepackage{amsmath}

\ifCLASSOPTIONcompsoc
 \usepackage[caption=false,font=normalsize,labelfont=sf,textfont=sf]{subfig}
\else
 \usepackage[caption=false,font=footnotesize]{subfig}
\fi
\usepackage[dvipsnames]{xcolor}
\usepackage{cleveref}
\usepackage{comment}
\usepackage{xspace}

\newcommand{\cvcv}{\textsc{cvc5}\xspace}
\newcommand{\cdclt}{CDCL$(T)$\xspace}

\newcommand{\define}[1]{\textsl{#1}}
\newcommand{\Mo}{\mathbf{I}}

\newcommand{\T}{\mathcal{T}}
\newcommand{\sbool}{\textsc{Bool}\xspace}

\begin{document}
%
\title{Partitioning Strategies for Distributed SMT Solving$^{\infty\!\!\!}$\thanks{$^{\infty}$The first author is supported by a U.S. Department of Energy Computational Science Graduate Fellowship under Award Number DE-SC0020347.  The work was also supported by an Amazon Research Award and the Stanford Center for Automated Reasoning.}}


%
\author{\IEEEauthorblockN{Amalee Wilson\IEEEauthorrefmark{1},
Andres Noetzli\IEEEauthorrefmark{2},
Andrew Reynolds\IEEEauthorrefmark{3}, 
Byron Cook\IEEEauthorrefmark{4},
Cesare Tinelli\IEEEauthorrefmark{3}, and
Clark Barrett\IEEEauthorrefmark{1}}
\IEEEauthorblockA{\IEEEauthorrefmark{1}Stanford University,
Stanford, USA \{amalee, barrett\}@cs.stanford.edu}
\IEEEauthorblockA{\IEEEauthorrefmark{2}Cubist, Inc., San Diego, USA n@cubist.dev}
\IEEEauthorblockA{\IEEEauthorrefmark{3}The University of Iowa, Iowa City, USA \{andrew-reynolds, cesare-tinelli\}@uiowa.edu}
\IEEEauthorblockA{\IEEEauthorrefmark{4}Amazon Web Services, Seattle, USA byron@amazon.com}}


\maketitle

\begin{abstract}
For many users of Satisfiability Modulo Theories~(SMT) solvers, the solver's performance is the main bottleneck in their application.  One promising approach for improving performance is to leverage the increasing availability of parallel and cloud computing.  However, despite many efforts, the best parallel approach to date consists of running a portfolio of solvers, meaning that performance is still limited by the best possible sequential performance.
In this paper, we revisit divide-and-conquer approaches to parallel SMT, in which a challenging problem is partitioned into several subproblems.  We introduce several new partitioning strategies and evaluate their performance, both alone as well as within portfolios, on a large set of difficult SMT benchmarks.
We show that hybrid portfolios that include our new strategies can significantly outperform traditional portfolios for parallel SMT.
\end{abstract}


%
\IEEEpeerreviewmaketitle

\section{Introduction} \label{intro}


State-of-the-art Satisfiability Modulo Theories (SMT) solvers such as Bitwuzla \cite{bitwuz}, \cvcv \cite{cvc}, MathSAT \cite{mathsat5}, OpenSMT \cite{opensmt}, Yices2 \cite{yices}, and Z3 \cite{z3} are widely used as reasoning engines in the context of verification~\cite{verification}, model checking~\cite{modelchecking}, security~\cite{security}, synthesis~\cite{synthesis}, test case generation~\cite{testcase}, scheduling~\cite{scheduling}, and optimization~\cite{optimization}.

For many users of SMT solvers, the solver's performance is a bottleneck for their application, and so improving solver performance continues to be a top priority for solver developers.
Today, most of the aforementioned solvers remain single-threaded, and performance improvements have primarily been achieved through new solving techniques and heuristics.
With the increasing availability of CPUs with large numbers of cores, high-performance computing (HPC), and cloud computing, a natural question is whether these resources together with parallel algorithms for SMT could be used to significantly boost solver performance.  Current research in this area can be divided into two main directions: \emph{portfolio solving} and \emph{divide-and-conquer solving}.

Portfolio solving is an approach in which multiple solvers (or different configurations of a solver), attempt to solve the same (or perturbed but equivalent) SMT problem in parallel \cite{concurrentZ3}.
It is well-known that SMT solvers are highly sensitive to small perturbations, which can dramatically improve or degrade their performance.  While efforts to reduce this sensitivity are an interesting research direction, it is difficult because of the inherent instability of the heuristics used and the uneven nature of the search space.  Portfolio solving aims to leverage this sensitivity by sampling from the possible configurations with the hope of finding one that performs well.  In fact, this strategy \emph{can} produce significant speed-ups and is currently considered the best way to leverage distributed computing resources to improve performance. 

Of course, naive portfolio solving is always limited by the best possible sequential performance, meaning that beyond some point, additional parallel resources do not help.  Also, portfolios are empirically ineffective for some classes of benchmarks.  For these reasons, it is appealing to pursue the alternative ``divide-and-conquer'' strategy.
In this approach, a problem is partitioned into independent subproblems in such a way that solving the subproblems provides a solution to the original problem.  The hope is that because the subproblems have smaller search spaces, solving them in parallel will be faster than solving the original problem.  Divide-and-conquer has the potential to dramatically outperform the best sequential performance, but only if an effective partitioning algorithm can be discovered.  Such an algorithm has been elusive.

One promising direction is to adapt the \textit{cube-and-conquer}~\cite{cubeAndConquer} approach that has been successfully applied to the more basic problem of Boolean satisfiability (SAT). 
In this approach, a partitioning heuristic is used to select a set of $n$ Boolean variables.  
Typically, a \emph{lookahead} heuristic~\cite{Heule2021LookAheadBS} is used, which chooses variables that, when assigned, most significantly prune the search space.
These variables are used to partition the problem into $2^n$ independent subproblems, which are then solved in parallel.  Unfortunately, attempts to adapt this approach for SMT
have had limited success.  In fact, for some cases, it has been observed that more partitioning is associated with larger, not smaller, runtimes~\cite{AnttiLookahead}. 

In this paper, we introduce several new partitioning strategies which build on---but also go beyond---the basic cube-and-conquer approach.  In particular, we look at different ways of combining \emph{sources} for collecting atoms (the SMT version of variables) with ways of using those atoms to create different \emph{partition types}.
We evaluate an implementation of these strategies in \cvcv on a diverse set of benchmarks from the SMT-COMP cloud track and from previous work on parallel SMT.
We show that a portfolio of partitioning strategies outperforms individual strategies, and we introduce the notion of a \emph{graduated} portfolio which performs particularly well.  We also show that \emph{hybrid} portfolios combining partitioning and traditional portfolio strategies perform even better. Finally, we show that using a \emph{multijob} scheduling algorithm for the partitioning portfolio accelerates performance even more. 
 We also demonstrate that these approaches scale, i.e., we continue to get additional speedup with more parallelism.

In summary, our contributions are the following:
\begin{itemize}
    \item the introduction of several novel partitioning strategies for parallel divide-and-conquer SMT solving;
    \item the introduction of graduated and hybrid partitioning portfolio strategies; and
    \item an implementation and evaluation of these strategies on a large set of benchmarks, including the first empirical results demonstrating a parallel solving technique that significantly outperforms a traditional non-partitioning portfolio and continues to do so as the number of partitions is increased.
\end{itemize}

\section{Preliminaries} \label{prelim}

We assume the standard many-sorted first-order logic setting with the usual
notions of signature, term, and interpretation.
A \emph{theory} is a pair $\T = (\Sigma, \Mo)$ where
$\Sigma$ is a signature and $\Mo$ is a class of $\Sigma$-interpretations.  For convenience, we assume a fixed background theory $\T$ with signature $\Sigma$ including the Boolean sort \sbool. We further assume that all terms
are $\Sigma$-terms, that entailment ($\models$) is
entailment modulo $\T$, equivalence is equivalence modulo $\T$, and that interpretations are $\T$-interpretations.  An \emph{atom} is a term of sort \sbool that does not contain any proper sub-terms of sort \sbool.  A \emph{literal} is either an atom or the negation of an atom.  A \emph{cube} is a conjunction of literals.
A formula $\varphi$ is a term of sort \sbool and is
\define{satisfiable} (resp., \define{unsatisfiable})
if it is satisfied by some (resp., no) interpretation in $\Mo$.  A formula whose negation is unsatisfiable is \define{valid}.

In this paper, we discuss partitioning algorithms that make use of the \cdclt framework employed by modern SMT solvers.
We give a brief overview of the \cdclt framework and introduce the relevant terminology.
Then, we describe how partitioning can be used for parallel SMT solving.

\medskip
\noindent
{\bf CDCL(\textit{T})-based SMT solvers}
solve problems via the cooperation of a SAT solver and one or more theory solvers~\cite{cdcl}. 
The role of the SAT solver in this framework is to build a truth assignment $M$ that satisfies the Boolean abstraction of the problem.  Typically, $M$ is built incrementally, and each time the SAT solver assigns a value to an atom, it calls the theory solvers to check whether $M$ is consistent with $\T$.  Theory solvers return conflict clauses and optionally new lemmas to the SAT solver. 
A conflict clause is a valid disjunction of literals that is falsified by $M$, and a lemma is any other heuristically-chosen valid formula.
When lemmas and conflict clauses are received by the SAT solver, they are added to the original problem. 
The process of finding satisfying assignments is repeated until one of two outcomes is achieved: either $M$ is a complete SAT assignment and no conflicts are detected by the theory solvers, meaning the problem is satisfiable; or, an unrecoverable conflict is derived, and the problem is therefore unsatisfiable.



\medskip
\noindent
{\bf Parallel SMT Solving with Partitioning}
The satisfiability of a formula $\phi$ can be determined in parallel by dividing it into $n$ independent \emph{subproblems} $\phi_{1}$, $\ldots$, $\phi_{n}$.  Provided the disjunction $\phi_{1} \vee \dots \vee \phi_{n}$ is equisatisfiable with $\phi$, if any of the subproblems are satisfiable, then the original problem is satisfiable, and if all of the subproblems are unsatisfiable, then the original problem is unsatisfiable. 
In this simple scenario, no synchronization is necessary during solving, because the subproblems are independent.

A \emph{partitioning strategy} constructs subproblems $\phi_{i}$ of the form  $\phi \land p_i$.  We call each $p_i$ a \emph{partitioning formula} and refer to each subproblem as a \emph{partition}.
%
%
Though not required for correctness, it is generally desirable for the partitions to be disjoint (i.e., for each $i\not=j$, the formula $\phi \land p_i \land p_j$ is unsatisfiable)  to avoid performing duplicate work. In the cube-and-conquer partitioning strategy, a set of $N$ atoms is selected, and each of the $2^N$ possible cubes using these atoms is used as a partitioning formula, resulting in $2^{N}$ partitions.
%
Scattering~\cite{SATScatter} is an alternative strategy which differs from cube-and-conquer in that it creates partitioning formulas that are not cubes.  Instead, scattering produces a series of $N$ partitioning formulas as follows.  The first partitioning formula is some cube $C_1$.  The second is $\neg(C_1) \wedge C_2$ for some new cube $C_2$.  The next is $\neg C_1 \wedge \neg C_2 \wedge C_3$ for a new cube $C_3$, and so on.  The $N^{th}$ partitioning formula is simply $\neg C_1 \wedge \cdots \wedge \neg C_{N-1}$.  Note that by construction, the partitioning formulas are disjoint.  However, there is considerable freedom in how the cubes are chosen.

\section{Related Work} \label{relwork}

As mentioned, much of the existing research literature on parallel and distributed SMT solving focuses on \emph{portfolios}.  A portfolio consists of multiple solver instances running in parallel, each of which attempts to solve the same problem.  The instance that finishes first produces the result and ends the portfolio run. Each instance differs from the others in some way: a different solver or configuration is used, or the problem has been perturbed in some equisatisfiable way.  Some portfolio frameworks enhance this basic strategy by sharing information (e.g., learned lemmas or clauses) among the solver instances running in parallel.
Z3 was one of the first solvers to support portfolio solving with sharing~\cite{concurrentZ3}.  SMTS~\cite{SMTS} is a parallel framework that also supports portfolio solving with sharing~\cite{clauseSharingCloud}.
In fact, SMTS implements the parallelization tree formalism~\cite{AnttiBookChapter} \cite{AnttiPartitioning2015}, which involves recursively combining both partitioning and portfolio solving.  
Our work mainly explores partitioning.  We focus on finding specific effective partitioning strategies, which could then be integrated into a framework such as SMTS. Interestingly, the two approaches (portfolio solving and partitioning) \emph{can} be effectively combined as observed in~\cite{SMTS} and as we discuss in Section~\ref{evaluation}.  Note that we do not (yet) consider information sharing, as this would add another layer of complexity, and there is enough to understand without it.  We consider the addition of information sharing a promising direction for future work.

There is also previous work on partitioning strategies.
\textit{Cube-and-conquer} has been successfully applied to SAT problems \cite{cubeAndConquer}, but there is no consensus on how this approach should be lifted to the SMT context. 
OpenSMT2 supports two different lookahead strategies for creating cubes in a cube-and-conquer-like partitioning strategy~\cite{AnttiLookahead}.  One is based on the global number of free atoms, and the other is based on the 
number of unassigned atoms in the clauses.  Previous evaluations using these strategies were mixed, with the strategies performing well on some benchmarks but not on others.
OpenSMT2 also supports a \emph{scattering} strategy that was originally developed for solving SAT problems~\cite{SATScatter}. 
In their implementation, each cube is obtained by taking atoms from the decisions made up along a particular search branch during a run of OpenSMT2.  The number of literals in each cube varies according to a heuristic.  Details can be found in~\cite{SATScatter,clauseSharingCloud,AnttiPartitioning2015}.
When the scattering partitioning strategy was compared to portfolio solving in~\cite{clauseSharingCloud}, they found that portfolio performed better on quantifier-free linear real arithmetic (QF\_LRA) benchmarks, especially on unsat problems.
We compare all of the OpenSMT2 strategies with our own strategies in our evaluation.

PBoolector~\cite{pboolector}, a parallel SMT solver built on top of the Boolector SMT solver~\cite{boolector}, uses a cube-and-conquer style strategy for QF\_BV SMT formulas, with the goal of evaluating whether lookahead methods
work well in combination with term-rewriting rules and bit-blasting techniques. 
On quantifier-free bitvector (QF\_BV) benchmarks, PBoolector saw familiar results: each configuration of their solver performed well on a subset of the benchmarks while performing poorly on others.  Because of our focus on partitioning strategies rather than implementations and the similarity of its partitioning strategy (lookahead-based) to that of OpenSMT2, we do not directly compare with PBoolector, though this would also be an interesting direction for future work.


Previous work on partitioning has also been limited in terms of which SMT-LIB logics were supported: benchmarks over quantifier-free uninterpreted functions (QF\_UF) were used in~\cite{AnttiPartitioning2015}, QF\_BV benchmarks in~\cite{pboolector}, and both QF\_UF and QF\_LRA benchmarks in~\cite{AnttiLookahead, SMTS, clauseSharingCloud}.
We are the first to implement a general-purpose partitioning strategy that works for all SMT-LIB logics. We comment on which types of problems are well-suited for our partitioning algorithms in Section \ref{evaluation}.

\section{Partitioning} \label{partitioning}
\newcommand{\atomSource}{\ensuremath{\mathit{atomSource}}\xspace}
\newcommand{\atomSelHeur}{\ensuremath{\mathit{atomSelHeur}}\xspace}
\newcommand{\cubeSize}{\ensuremath{\mathit{cubeSize}}\xspace}
\newcommand{\makePartitions}{\ensuremath{\mathit{makePartitions}}\xspace}
\newcommand{\tHeur}{\ensuremath{\mathit{tHeur}}\xspace}
\newcommand{\isTimeToPartition}{\ensuremath{\mathit{isTimeToPartition}}\xspace}
\newcommand{\collectAtoms}{\ensuremath{\mathit{collectAtoms}}\xspace}
\newcommand{\lits}{\ensuremath{\mathit{atoms}}\xspace}
\newcommand{\ptype}{\ensuremath{\mathit{ptype}}\xspace}
\newcommand{\timeToPart}{\ensuremath{\mathit{timeToPart}}\xspace}


In this section, we introduce a set of partitioning strategies parameterized in four dimensions: atom source, selection heuristic, partition type, and partition timing. 
Pseudocode for partitioning based on these parameters is given in \Cref{alg:cap}. 
More details on these parameters and their relationship to Algorithm 1 are given in the following subsections, but briefly, the atom source and selection heuristic specify \textit{what} the partitions are made of, the partition type specifies \textit{how} the partitions are made, and the partition timing specifies \textit{when} the partitions are made.

In all cases, partitions are made by invoking an instrumented version of an SMT solver that calls \Cref{alg:cap} periodically during the solving process.  We call this solver the \emph{partitioning solver}.
By instrumenting an existing solver, our approach takes advantage of well-tested infra\-structure for parsing, preprocessing, and reasoning about SMT problems.
The first step in Algorithm 1 is to check whether, according to the \emph{partition timing} heuristic (see Section~\ref{partition_timing}), it is the right time to make a partition.  If not, nothing is done.  Otherwise, a set of atoms is collected from the specified source (see Section~\ref{atom_selection}).  If an insufficent number of atoms is collected, again, nothing is done. Otherwise, \makePartitions is called.  Depending on the \emph{partition type} (see Section~\ref{partition_type}), we may be done or we may need to continue.  \makePartitions returns true if partitioning is done.
Whenever \Cref{alg:cap} yields a new partition, but partitioning is not yet done, the partitioning solver blocks the part of the search space corresponding to the generated partitioning formula by adding the negation of the partitioning formula as a lemma (called a \emph{blocking lemma}).
The partitioning solver then continues to work on solving the problem until its next call to Algorithim 1.

It is possible that the partitioning solver actually solves the problem.
If the solver determines that the problem is satisfiable, or if it finds that the problem is unsatisfiable before any partitions have been made, then the problem has been solved, and there is no need to continue or to solve any partitioned formulas.
However, if the partitioning solver returns unsatisfiable after having emitted some partitions, then these partitions still need to be solved.  This is because of the blocking lemmas that were added which prune the parts of the search space (in the partitioning solver) covered by the emitted partitions.


\algrenewcommand\algorithmicrequire{\textbf{Input:}}
\algrenewcommand\algorithmicensure{\textbf{Output:}}
\begin{algorithm}
\caption{Partitioning strategy pseudocode.}\label{alg:cap}
\begin{algorithmic}[1]

\Require $N \geq 2$, $\cubeSize = \log_{2}(N)$
\Require $\atomSource \in \{\text{HEAP, DECISION, CL}\}$
\Require $\atomSelHeur \in \{\text{RAND, SPEC}\}$
\Require $ptype \in \{\text{CUBE, SCATTER}\}$
\Require $t_1, t_2 \geq 0, \tHeur \in \{\text{CHECK, TIME}\}$
\Ensure returns true iff done partitioning
\State $\timeToPart \gets \isTimeToPartition(t_1, t_2, \tHeur)$
\If{$\timeToPart$}
    \State $ \lits \gets \collectAtoms(\atomSource, \atomSelHeur) $
    \If{$\lits\mathit{.size()} \geq \cubeSize$}
        \State $\lits\mathit{.resize}(\cubeSize)$
        \If{$\makePartitions(\ptype, \lits, N)$}
            \State \Return true
        \EndIf
    \EndIf
\EndIf
\State \Return false
\end{algorithmic}
\end{algorithm}

\subsection{Atom Source and Selection Heuristics} \label{atom_selection}
There are two parameters for atom selection: $\atomSource$ and $\atomSelHeur$. The $\atomSource$ parameter describes where the atoms should come from. 
We explore three different sources of atoms for our partitioning strategies: the SAT heap, which is a priority queue of Boolean variables in the SAT solver; the decision trail of the SAT solver, which contains the decisions made by the SAT solver along its current search branch; 
and conflict-or-lemma (CL) atoms, which are atoms contained in the lemmas and conflict clauses sent from the theory solver to the SAT solver. These sources of atoms correspond to HEAP, DECISION, and CL in \Cref{alg:cap}, respectively.
The $\atomSelHeur$ parameter describes how atoms should be selected from the available atoms in the source. 
Every source supports selecting atoms at random (RAND in \Cref{alg:cap}). The other option is to use a heuristic specific to the source (SPEC). We describe these below.
To guarantee that partitions can be made quickly, each of the heuristics is lightweight and does not rely on sophisticated or computationally expensive score calculations.

As mentioned above, when \atomSource is HEAP, the source of atoms is an internal data structure in the SAT solver.  Typically, SAT solvers make decisions based on the \emph{activity} score of each variable.  The activity of a variable is determined by how often it appears in conflicts~\cite{var_scoring}, and the variable with the highest score is used when the SAT solver makes decisions. 
When using HEAP as its source, the SPEC heuristic simply chooses the \cubeSize variables with the highest activity scores.   The rationale is that highly active variables may be a good choice for helping shape partitions.
%
Note that this heuristic requires no additional computation by the partitioning algorithm because the SAT solver already orders the variables by their activity.

Variables that are good for SAT decisions may not always be ideal for SMT partitions.  For example, some high-activity variables may be closely related to other high-activity variables, because they represent theory atoms that contain similar terms.  Ideally, however, variables used in partitions should be as \emph{independent} as possible, so that each partition has roughly the same difficulty.
The DECISION option attempts to address this weakness.  It uses the decision trail as a source of atoms.  The decision trail contains all the variables that have been decided on along a particular branch of the search tree during the solving process.  The rationale is that variables in the decision trail are, roughly speaking, more likely to be independent (for example, if two variables entail each other, then deciding on one will always propagate the other, so they cannot both be in the decision trail at the same time).
When selecting atoms from the decision trail, the SPEC heuristic chooses the earliest decisions in the trail.
As before, this heuristic requires no additional computation by the partitioning algorithm, because the decisions are stored in the trail from least to most recent.

Finally, the CL option uses conflict clauses and lemmas coming from theory solvers as the atom source.  Intuitively, atoms that appear in conflict clauses and lemmas are those that are ``contributing'' in some way to the solution process.  While the appearance of an atom in a conflict clause is reflected in its activity score, an appearance in a lemma has no effect on the activity score.  This is because when lemmas are generated, they are simply added as additional clauses.  The role of a lemma is to help guide the search in a theory-specific way.  Thus, we expect that atoms appearing in lemmas may be important.
When selecting atoms from conflict clauses and lemmas, the SPEC heuristic selects those atoms that occur most frequently. These atoms are tracked as they are sent from the theory solver to the SAT solver, and a counter is maintained for each atom. 


There are two additional issues to consider when selecting atoms: the number of atoms to use for each partition and filtering out unusable atoms. In Algorithm 1, we fix the number of atoms per partition, $\cubeSize$, to be $log_{2}(N)$ where $N$ is the number of requested partitions. However, as we discuss below, for the SCATTER partition type, this is not necessary and \cubeSize could be set as an additional parameter. Regarding filtering, because the construction of partitions is done by appending a partitioning formula to the original formula, anything appearing in the partitioning formula must make sense in this context.  In particular, if an atom contains some symbol generated internally by the solver (i.e., not appearing in the original problem), we filter that atom out.  This filtering is done during the $\collectAtoms$ routine. 
 

\subsection{Partition Type} \label{partition_type}
We consider two ways of creating partitioning formulas from the selected atoms.  The first, the CUBE partition type, follows the \textit{cube-and-conquer} approach. The second, the SCATTER partition type, uses a scattering strategy.

\subsubsection{Cubes}
The cube strategy requires $log_{2}(N)$ atoms to be selected during the atom selection phase. Once the required number of atoms has been collected, they are immediately used to create $N$ mutually exclusive cubes.  The partitioning solver then terminates.
These cubes, $C_{1}$ through $C_{2^n}$, correspond to each possible conjunction of atoms that can be created from the selected atoms. 
For example, if the two atoms selected are $x_1$ and $x_2$, then the following partitions would be emitted: $C_{1} = x_1 \wedge x_2$,  $C_{2} = \neg x_1 \wedge x_2$, $C_{3} = x_1 \wedge \neg x_2$, and $C_{4} = \neg x_1 \wedge \neg x_2$. 

\subsubsection{Scattering}
Scattering is a dynamic strategy for creating partitioning formulas. 
When the partition type is SCATTER, Algorithm 1 produces only a single partition with each call to \makePartitions.  The partitioning formula constructed at each call takes the current cube and conjoins it with the negation of previous cubes, as described in Section~\ref{prelim}.  After each generated partition, the negation of the partitioning formula is added as a lemma to the partitioning solver, to ensure that it explores a different part of the search space during the rest of its run.

Note that \cubeSize does not have to be equal to $\log_{2}(N)$ when using scattering.  Indeed, it can even vary from partition to partition.  In~\cite{AnttiPartitioning2015}, a particular strategy is suggested that does vary the \cubeSize across partitions.  In this paper, we simply use $\log_{2}(N)$, for the \cubeSize, but a systematic exploration of this parameter for the SCATTER partition type is an interesting direction for future work.

Note that it takes at least $N-1$ calls to Algorithm 1 to compute $N$ partitions with the SCATTER partition type. The final partition is emitted immediately after the $N-1^{st}$ partition, because the final partition is simply the negation of all previously used cubes. 

\subsection{Partition Timing} \label{partition_timing}

Partition timing specifies \emph{when} partitions should be made. 
There is a trade-off between collecting sufficient information to create good partitions and avoiding spending unnecessary time on partitioning that could have been used for solving.
Algorithm 1 supports two different kinds of partition timing. 
The first, when \tHeur is CHECK, simply counts the number of times that Algorithm 1 has been called (in our implementation, this is done during the \emph{check} phase as we explain in Section~\ref{evaluation} below). The second possibility is TIME, which simply measures the amount of time that has passed. 
In Algorithm 1, two inputs, $t_1$ and $t_2$, are used for timing.
The $t_1$ parameter determines how long to wait (either number of checks or time in seconds) before the partitioning solver creates any partitions.  The idea of this parameter is to allow the partitioning solver to make some progress and get into an interesting state before starting partitioning.  The $t_2$ parameter determines how long (again in either checks or seconds) to wait \emph{between} each pair of partitions.
Note that for the CUBE partition type, $t_2$ is irrelevant, because all partitions are created at once.

There is another trade-off between predictability of partitioning time and predictability of partitioning formulas. 
When counting checks, the partitioning formulas are deterministic (as long as the execution of the SMT solver is deterministic).
When using time, however, there can be variation in the current state of the solver at time $t$ from one run the to next. 
Thus, it may seem like check counting is preferable.  The problem is that the number of checks varies greatly from one problem to the next.  One SMT formula may trigger thousands of checks in the solver in the first minute of solving, while other SMT formulas have only a handful. 
Thus, for predictable partitioning time (though less predictable partitions), it can be better to use time instead of check counting to help ensure that the time to create partitions is relatively stable across many problems.  We discuss this further in Section~\ref{selection}, below.

\section{Evaluation} \label{evaluation}

We instrumented \cvcv to be a partitioning solver by (i) implementing Algorithm 1 in \cvcv; and (ii) having \cvcv call Algorithm 1 after each decision made by its internal SAT solver.  Below, we report on several sets of experiments with this instrumented version of \cvcv.
We ran all experiments reported here on a cluster with 26 nodes running Ubuntu 20.04 LTS, each with 128 GB of RAM, and two Intel Xeon CPU E5-2620 v4 CPUs with 8 cores per CPU. 

In our evaluation, we compare several configurations of our \cvcv-based partitioning solver and an OpenSMT2-based partitioning solver on a set of benchmarks drawn from the cloud track of the 2022 edition of SMT-COMP~\cite{smtcomp2022}, the SMT-LIB QF\_LRA benchmarks, and the SMT-LIB QF\_UF benchmarks~\cite{smtlib}.  The SMT-COMP cloud benchmarks were selected by the SMT-COMP organizers to be challenge problems for parallel solving and are thus a good target for this work. 
The QF\_LRA and QF\_UF benchmarks have been the subject of previous studies on parallel solving~\cite{AnttiLookahead,AnttiPartitioning2015,clauseSharingCloud}.
We exclude benchmarks based on several criteria.  First, we exclude any benchmark with quantifiers.  The challenges for quantified benchmarks are typically the result of too many possible quantifier instantiations.  In contrast, partitioning targets challenges stemming from large Boolean search spaces.  Studying the interaction of these two challenges is an interesting direction for future work but is beyond the scope of this paper.
Second, we exclude benchmarks that are solved in less than 600 seconds by the sequential version of \cvcv.  This is simply to focus on problems that are challenging for sequential solvers.
Third, if no partitions can be made by the partitioning solver, the benchmark is excluded.  This can happen if the SAT solver makes no or almost no decisions and Algorithm 1 is not called enough times to create partitions.
We consider such benchmarks poor candidates for solving via partitioning.
Finally, benchmarks that are solved during partitioning by any partitioning solver are excluded, again because we consider them easy, and furthermore, our aim is to compare partitioning strategies, not partitioning solvers.
After these filters are applied, we are left with 214 challenging benchmarks in 5 SMT-LIB logics: QF\_LRA (139), QF\_IDL (48), QF\_LIA (16), QF\_UF (7), and QF\_RDL (4). 

To measure the performance of a particular partitioning strategy on a given benchmark, we first measure the time to partition it. Then, we run each partition on the cluster (with a maximum of 16 jobs per node, i.e., one per core, and 8 GB of memory per job) using the \cvcv SMT-COMP~2022 script (which runs a theory-dependent sequential instance of \cvcv) and record the time.
Then, we record the total solving time for the benchmark as the sum of the partitioning time and either the maximum time required to solve any of its partitions, if the benchmark is unsatisfiable, or the minimum runtime of the satisfiable partitions, if the benchmark is satisfiable. Note that this simulates a parallel run in which all of the partitions are run simultaneously on separate cores.

To evaluate a partitioning strategy on the set of benchmarks as a whole, we use the PAR-2 score~\cite{satcomp}, which is also used for scoring the annual SAT competition. The PAR-2 score is the sum of the runtimes for all solved instances plus \emph{twice} the timeout value multiplied by the number of unsolved instances. The lower the PAR-2 score is, the better. We use the PAR-2 score because it provides a single metric that incorporates both runtime and number of benchmarks solved.
In this work, we are primarily interested in the scalability of the different partitioning strategies, so most experiments measure the PAR-2 score for different numbers of partitions.
We use a 20 minute timeout when solving partitions, which includes both the time to partition and the time to solve.


\subsection{Partitioning Strategies} \label{selection}

\begin{table}[!t]
\renewcommand{\arraystretch}{1.2}
\caption{Effect of Wait Times on Cubing of 125 QF\_LRA benchmarks}
\label{ptime_table}
\centering
\begin{tabular}{|c|c|c|}
\hline
Time & Solved & PAR-2\\
\hline
1s & 40 & 223636\\
\hline
3s & 42 & 219187\\
\hline
15s & 42 & 219313\\
\hline
\end{tabular}
\end{table}


We first explore the different partitioning strategies in the parameter space of Algorithm 1, starting with the partition timing parameter.

First, we measure the impact of $t_1$ when using TIME for \tHeur.
In general, we find that if the value of $t_1$ is too small, then the partitioning solver does not have enough time to provide good atoms, but beyond a threshold there is little additional benefit.
To demonstrate this, Table~\ref{ptime_table} shows a set of results using different values of $t_1$. 
For these runs, we set \atomSource to HEAP, \ptype to CUBE, and \atomSelHeur to SPEC, and run on a subset of benchmarks consisting of 125 QF\_LRA benchmarks.
We see that 
with $t_1=3$, we are able to
solve more problems and obtain a better PAR-2 score than 
with $t_1=1$.
At the same time, increasing $t_1$ to $15$ does not result in any additional problems being solved and has a small negative effect on the PAR-2 score. 
Additional experiments (not shown here) using other parameter settings and benchmarks show similar results.  Based on these observations, we use $t_1=3$ when using TIME.

\begin{figure*}[!t]
\centering
\subfloat[Decision Scattering]{\includegraphics[width=.3\textwidth]{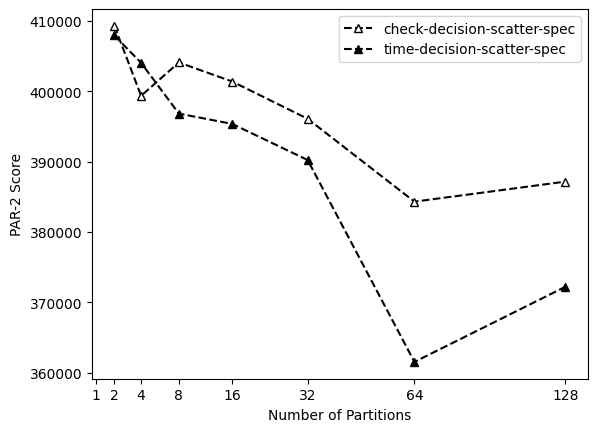}
\label{time_v_check_decision}}
\hfil
\subfloat[Heap Scattering]{\includegraphics[width=.3\textwidth]{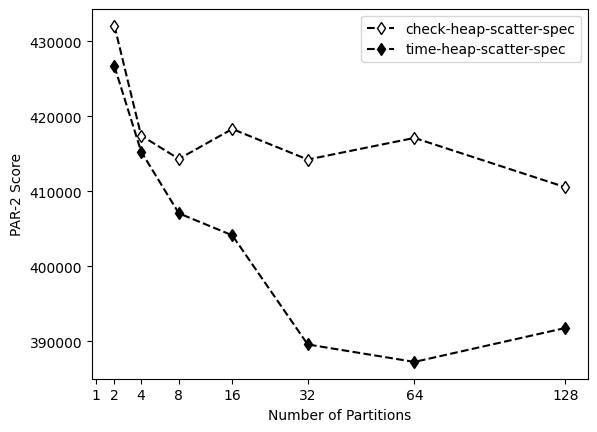}%
\label{time_v_check_heap}}
\hfil
\subfloat[CL Scattering]{\includegraphics[width=.3\textwidth]{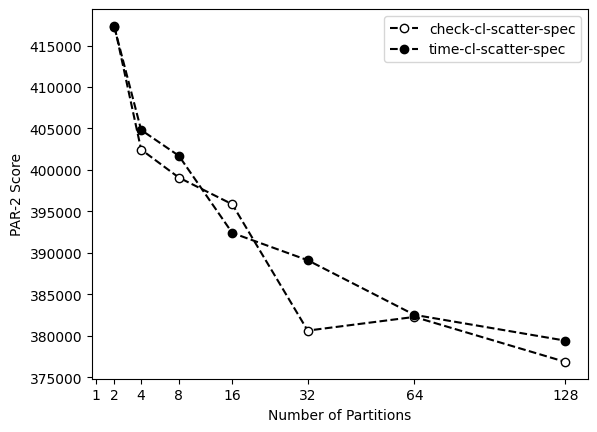}%
\hfil
\label{time_v_check_lemma}}
\caption{Comparison of using TIME vs CHECK for \tHeur when scattering.}
\label{fig_time_v_check}
\end{figure*}

Next, we compare $\tHeur = \text{CHECK}$ and $\tHeur = \text{TIME}$.
For these experiments, we use \ptype= SCATTER (since otherwise $t_2$ plays no role) and \atomSelHeur= SPEC.
For \tHeur= TIME, we use $t_1=3$ and $t_2=0.1$, so the first call to Algorithm 1 is after three seconds, and each subsequent call is after an additional tenth of a second. 
For the CHECK runs, we use $t_1=1$ and $t_2=1$, so the first call is after the first check, and each subsequent call is after the next check. 
Figure \ref{fig_time_v_check} shows the results\footnote{These and subsequent experiments are on the full set of 214 benchmarks.} of our experiments for different atom sources and partition sizes.
Figures \ref{time_v_check_decision} and \ref{time_v_check_heap} show that when $\atomSource$ is DECISION or HEAP,
TIME clearly outperforms CHECK. 
With CL, the results are not conclusive, with both strategies performing similarly.

One likely reason for the poor performance of CHECK in the first two cases is that with $t_1=1$ we start partitioning on the first call to Algorithm 1.  As we mentioned above, it is usually better to wait some time before partitioning.  However, there is a tremendous amount of variation in the
number of calls to Algorithm 1 across different benchmarks.  This makes it difficult to find values other than $1$ for $t_1$ and $t_2$ that work consistently across all benchmarks.  On the other hand, using TIME with $t_1=3$ and $t_2=.1$ tends to perform well across different benchmarks and parameters.  For these reasons, we use these parameters in the remaining experiments.

It is worth noting that the disadvantage of nondeterminism when using TIME remains.  In future work, we hope to find a way to get the consistency of TIME while also having deterministic results.


\begin{figure}[!t]
\centering
\includegraphics[width=0.3\textwidth]{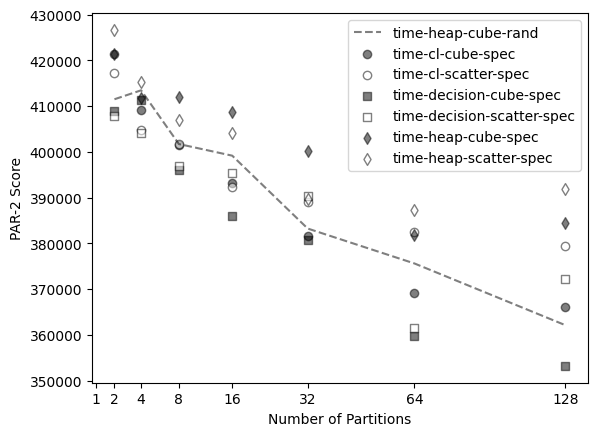}
\caption{Various partitioning strategies vs a random strategy.}
\label{vs_random}
\end{figure}

We now turn our attention to the other parameters of Algorithm 1.
Figure \ref{vs_random} shows the six possible strategies using \atomSelHeur= SPEC compared with a random strategy (TIME-HEAP-CUBE-SPEC) as a baseline (other random strategies are similar).
The HEAP strategies consistently perform worse than the random strategy, which suggests that relying on the SAT solver's priority queue heuristic is not particularly effective for making partitions. 

The DECISION strategies, on the other hand, both perform quite well, beating the random strategy most of the time.
This suggests that selecting the earliest decisions from the decision trail is a useful heuristic when selecting atoms for partitioning.
The CL strategies are mixed.  The CUBE variant performs favorably compared to random, while the SCATTER variant generally does not.
Based on these results, we choose three configurations as our top partitioning strategies, and refer to them in the following as decision-cube (TIME-DECISION-CUBE-SPEC), decision-scatter (TIME-DECISION-SCATTER-SPEC), and cl-cube (TIME-CL-CUBE-SPEC).

Clearly, there are many more possible strategies and variants that could be explored.  We expect this to be a fruitful direction for future work.

\subsection{Comparison to OpenSMT2 Partitioning Strategies} \label{comparison}

\begin{figure}[!t]
\centering
\includegraphics[width=0.3\textwidth]{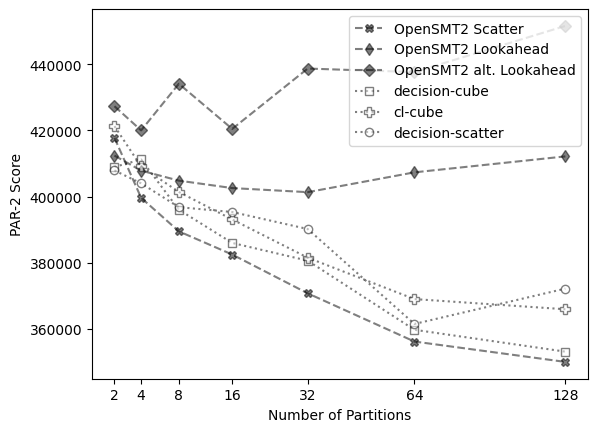}
\caption{\cvcv partitioning strategies vs OpenSMT2 partitioning strategies}
\label{vs_osmt}
\end{figure}

The most extensive studies on partitioning strategies in previous work are from the OpenSMT2 team.
We next compare our best partitioning strategies, decision-cube, cl-cube, and decision-scatter, to the three partitioning strategies available in OpenSMT2. 
Recall that none of the selected benchmarks are solved during partitioning, so this is a comparison only of the partitioning strategies, not the solvers. 
Figure \ref{vs_osmt} shows the results of this comparison. 
The two lookahead partitioning strategies in OpenSMT2 perform worse than our three best strategies.  This is partly because they are slow and often time out during partitioning. On the other hand, the OpenSMT2 scattering strategy performs very well. 
In particular, their scattering strategy outperforms our individual partitioning strategies, though decision-cube is quite close. Interestingly, the OpenSMT2 scattering strategy also uses decisions as its source of atoms, making it very similar to our decision-scatter strategy.  However, they have a few additional parameters that have been fine-tuned.  For example, they vary the number of literals per partitioning formula, something that our strategy does not do.  This suggests that our decision-scatter strategy could likely be improved in a similar way.
In the spirit of ``if you can't beat them, join them,'' we replace our decision-scatter strategy with the OpenSMT2 scattering strategy in our list of top performing strategies going forward and refer to it as osmt-scatter.


\subsection{Partitioning Portfolios}
\label{part-port}


\begin{figure*}[!t]
\centering
\subfloat[Partitioning Portfolios]{\includegraphics[width=.49\columnwidth]{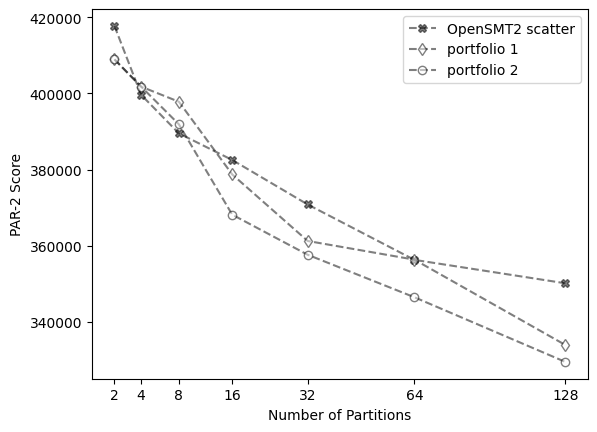}
\label{portfolio_vs_osmt}}
\hfil
\subfloat[Graduated Portfolio]{\includegraphics[width=.49\columnwidth]{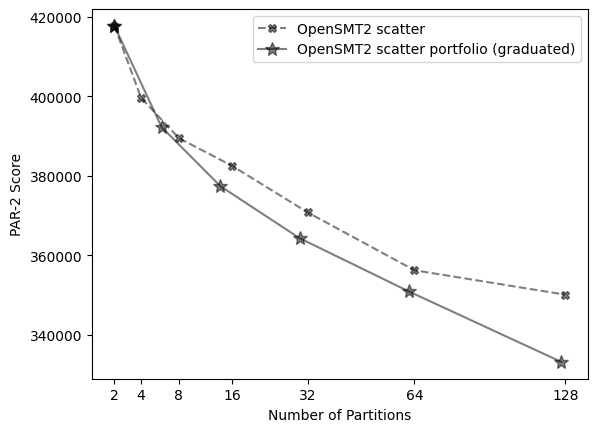}%
\label{iterative_portfolio}}
\hfil
\subfloat[Graduated Portfolio Comparisons]{\includegraphics[width=.49\columnwidth]{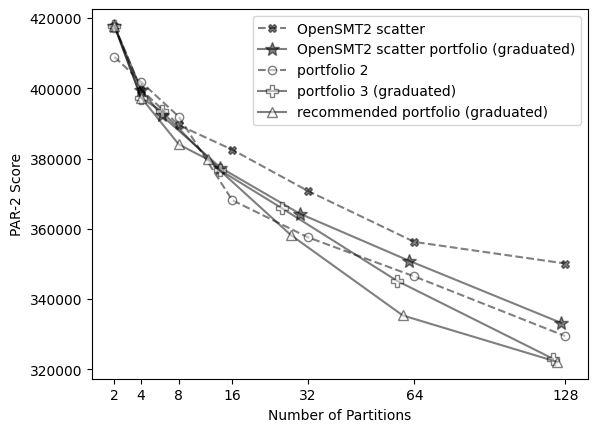}%
\hfil
\label{iterative_portfolio_options}}
\caption{Comparison of various partitioning portfolios.
}\label{portfolio_comparisons}
\end{figure*}

One consistent observation in this and previous work is that there is a lot of variation in how well strategies work across benchmarks.  Every strategy fails on some benchmarks and works well on others.  Given the success of portfolio solving in general, a natural question is whether a portfolio of partitioning strategies can outperform individual partitioning strategies.  It is not immediately obvious whether this should be true, since, to be fair, we require that partitioning portfolios divide up their partitioning budget.  For example, we would compare a partitioning portfolio using 2 strategies, each creating 16 partitions to an individual strategy that can use 32 partitions.


Figure~\ref{portfolio_vs_osmt} shows the results of using two different partitioning portfolios and compares them to the best individual strategy, osmt-scatter.
Portfolio 1 creates $N/2$ partitions with each of decision-cube and cl-cube and demonstrates that a portfolio of two (worse) strategies can do better than the best individual strategy. 
Portfolio 2 creates $N/2$ partitions with decision-cube, $N/4$ partitions with cl-cube, and $N/4$ partitions with osmt-scatter and demonstrates that adding more variety in the portfolio can result in even better performance. 

These results suggest that the orthogonality of the individual strategies is high and that adding more strategies at the expense of the number of partitions can be beneficial.  To take this to its logical conclusion, we introduce what we call a \emph{graduated partitioning portfolio}, which maximizes the number of strategies included.
%
A graduated partitioning portfolio based on $m$ individual partitioning strategies
is constructed as follows.  First, take each of the $m$ individual strategies and parameterize them by $N$ for values of $N$ that are powers of 2, e.g., osmt-scatter-2 is the strategy that uses osmt-scatter to make 2 partitions, and cl-cube-64 is the strategy that uses cl-cube to make 64 partitions.  Now, rank all of the strategies, with smaller values of $N$ being ranked higher.  To break ties when $m>1$, use a ranking on individual strategies (for our top performing strategies, we rank osmt-scatter-$N$ higher than decision-cube-$N$ higher than cl-cube-$N$).  Finally, to obtain the graduated portfolio strategy for $N$ partitions, simply collect strategies from this list, in order, until it is no longer possible to add strategies without exceeding $N$.  For example, for $N=32$ and $m=3$, we would choose all of the 2-partition strategies, all of the 4-partition strategies, and the osmt-scatter-8 strategy, for a total of 26 partitions.  We do not attempt to use the remaining 6 partitions in our ``partition budget.''

To visualize the value of graduated partitioning, Figure \ref{iterative_portfolio} compares the stand-alone osmt-scatter strategy with a graduated partitioning portfolio version of the same strategy (i.e., $m=1$ and the only strategy is osmt-scatter).\footnote{To see the difference, note that for $N=16$, the stand-alone strategy runs osmt-scatter once to make 16 partitions, whereas the graduated portfolio runs osmt-scatter three times, making 2, 4, and 8 partitions, respectively.}
Notice that although the graduated portfolio uses two fewer partitions for every plotted value of $N$, it clearly outperforms the stand-alone strategy, especially as the number of partitions is increased. 

The next step is to see what happens when we use more than one strategy in the graduated partitioning portfolio. We experimented with all possible combinations of $m=1,2,3$ and our top performing strategies. Figure \ref{iterative_portfolio_options} shows selected results (and also the best strategies from Figures~\ref{vs_osmt}, \ref{portfolio_vs_osmt} and \ref{iterative_portfolio} for comparison).  Portfolio 3 is the graduated partitioning portfolio with $m=3$ using all three top-performing strategies: osmt-scatter, decision-cube, and cl-cube. 
Portfolio 3 outperforms portfolio 2 for large numbers of cores. 
However, the strongest portfolio uses $m=2$ and only combines osmt-scatter and decision-cube (in this case, it appears that the diversity of cl-cube does not compensate for the lack of additional versions of the other two strategies).  We refer to this as the ``recommended portfolio.''
Our recommended portfolio strategy has consistently better performance  than all other strategies we have considered, even though it uses fewer partitions than non-graduated strategies.

\subsection{Hybrid Portfolios}

Finally, we conclude by trying to address a very practical question.  Given our results, how should one go about using $N$-way parallelism to solve a challenging SMT problem?  In particular, how do the best partitioning approaches compare with traditional portfolio approaches.  To represent the latter, we use a \emph{scrambling} portfolio.  Given a problem and a value of $N$, we construct a scrambling portfolio of size $N$ by running $N$ versions of the problem: the original problem plus $N-1$ copies obtained using the SMT-COMP scrambler~\cite{smtscrambler} with different random seeds.  The runtime for the scrambling portfolio on the problem is then the minimum of these runtimes (we include the time required to scramble in the individual runtimes, but this is typically negligible).

\begin{figure}[!t]
\centering
\includegraphics[width=0.3\textwidth]{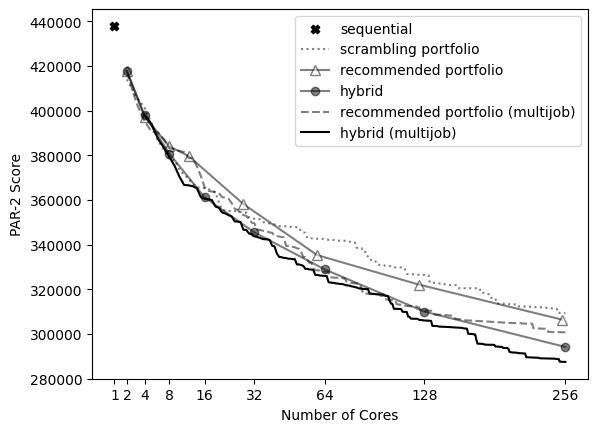}
\caption{Comparison of different portfolios. Hybrid and recommended portfolios are graduated.}
\label{vs_scrambling}
\end{figure}

Figure~\ref{vs_scrambling} compares, for different values of $N$, a scrambling portfolio with our recommended partitioning portfolio, based on a simulation where each partition or scramble is run on a separate core in parallel (it also includes the performance of a single sequential run for comparison).  
We see that the scrambling portfolio does indeed outperform the recommended partitioning portfolio for $N<32$, but the recommended partitioning portfolio wins as the number of cores gets larger.

However, continuing with our theme of ``if you can't beat them join them,'' we next consider a \emph{hybrid} portfolio that combines scrambling and partitioning.
For this approach, given $N$ cores, we simply run a scrambling portfolio of size $N/2$ and a recommended partitioning portfolio of size $N/2$.  The result is shown as ``hybrid'' in Figure~\ref{vs_scrambling}. 
Remarkably, the hybrid strategy clearly outperforms the scrambling portfolio, even for small numbers of cores.
We note that, to our knowledge, this is the first time any parallel strategy has been shown to be consistently superior to a traditional portfolio.

Finally, there is one more optimization we can apply: we can use the \emph{multijob} strategy referenced in the original cube-and-conquer paper~\cite{cubeAndConquer}.  The key idea is that because partitions can be very uneven (in runtime), using one core per partition means that many cores (the ones that get assigned easy partitions) will be idle most of the time.  The multijob strategy simply consists of using \emph{many more partitions} than cores, and then scheduling the partitions as cores become available.  This automatically load-balances the partitions and results in being able to include the results from more partitions without much additional (wall-clock) runtime.

The recommended portfolio (multijob) line shows the result of simulating a recommended portfolio that includes \emph{all} versions of osmt-scatter and decision-cube (from 2 up to 128)  for different numbers of cores.  We schedule the smaller partitions first, and no core is allowed to do more than 20 minutes of work.
%
The hybrid (multijob) shows the results of using $N/2$ cores for a scrambling portfolio and running the multijob scheduling algorithm on the other $N/2$ cores.
Note that the multijob approach cannot be used to accelerate scrambling portfolios, only partitioning portfolios.
The best strategy is hybrid (multijob) and with 256 cores, it improves the PAR-2 score by 34\% (compared to a single sequential solver).

\section{Conclusion} \label{conclusion}
We have shown that a portfolio of partitioning strategies, hybrid portfolios in particular, can outperform traditional portfolio solving for a range of SMT problems. 
These new strategies are a step towards 
better utilization of HPC and cloud systems for solving SMT problems.
Plans for future work include
improving our partitioning strategies, implementing information sharing, and exploring recursive partitioning, 
\bibliographystyle{IEEEtran}
\bibliography{paper}





\end{document}